\documentclass[12pt]{article}
\usepackage{graphics}
\begin{document} 
 
\title{Dynamical Nonsupersymmetry Breaking\footnote{Talk given at the
International Workshop on Perspectives of Strong Coupling Gauge
Theories (SCGT96), November 1996, Nagoya, Japan.}} 
 
\author{\normalsize B. Holdom\thanks{holdom@utcc.utoronto.ca}\\
\small {\em Department of Physics,}
\small {\em University of Toronto}\\
\small {\em Toronto, Ontario,}
\small M5S1A7, CANADA}
\date{}
\maketitle
\begin{picture}(0,0)(0,0)
\put(310,205){UTPT-97-02}
\put(310,190){hep-ph/9702250}
\end{picture}
\begin{abstract}
We emphasize the role that anomalous power-law
scaling of 4-fermion operators, occurring in the presence of new strong
interactions, could have in the generation of quark and lepton masses.
\end{abstract}
\vspace{4ex}

\section{Introduction}

Recent advances in the understanding of strong dynamics and
symmetry breaking in supersymmetric theories, as reported at
this conference, have made clear an impressive diversity of
phenomena which can arise in strongly interacting theories. This
has led to renewed efforts in the search for a plausible dynamical
theory of quark and lepton masses, especially in theories
incorporating supersymmetry. But the wide range of phenomena
uncovered in strongly interacting supersymmetric theories might
be considered to suggest a corresponding range of phenomena in
strong gauge dynamics in general. In particular one might expect
that the diversity of the phenomena would be at least as great
when the constraints of supersymmetry are removed. This
possibility is somewhat at odds with the typical viewpoint taken
in attempts to incorporate dynamical symmetry breaking, without
supersymmetry, into realistic models. Much of the
previous work over the years have focused on the construction of
models with dynamics resembling that of QCD as closely as possible.

In this talk we will illustrate the role that
non-QCD-like dynamics, in particular dynamics which gives rise
anomalous power-law scaling, could play in realistic models of
fermion mass. We describe how a model with a simple symmetry
structure can give rise to a wide variety of 4-fermion operators,
which in the presence of anomalous scaling, can give rise to a
mass spectrum with nontrivial, and potentially realistic, mixings
and hierarchies.

There are other reasons why a fresh perspective on
dynamical symmetry breaking may be useful. We begin by listing
some statements and implicit assumptions sometimes made
in connection with dynamical electroweak symmetry breaking,
and suggest why these statements may not be completely correct. 

\begin{itemize}
\item ``The top mass is so large that it must be involved
with the electroweak symmetry breaking dynamics''.

--- Not necessarily, since 175 is small compared to 1000. If a
dynamical fermion mass is responsible for the \(W\) and \(Z\)
masses, then we expect that fermion mass to be closer to
1000 GeV than to 175 GeV.
\item ``If the \(t\) mass has a dynamical origin then $\delta
\rho $ is too large.''

--- Not necessarily, if contributions to $\delta\rho $ are suppressed by
powers of 175/1000.
\item ``The dynamical origin of the \(t\) mass is distinct from that of
other quarks.'' For example in ``topcolor''\cite{b1} models it is
postulated that the top mass comes from
$\overline{t}t\overline{t}t$ operators rather than
$\overline{H}H\overline{t}t$ operators. Here \(H\) is the
TeV-mass fermion responsible for electroweak symmetry
breaking.

--- Not necessarily, since $\overline{H}H\overline{t}t$ operators
are by themselves still okay; we will see that these operators
make contributions to
$\delta\rho $ which are suppressed by powers of
175/1000. The real problem is somewhat removed from \(t\)
mass generation and has to do with is explaining why
isospin-violating operators of the form
$\overline{H}H\overline{H}H$ are not too large. We will argue
that this problem may be resolved in a situation where some
operators enjoy more anomalous scaling than other operators.
\item ``The scale of new flavor physics must be well above a TeV.''

--- Not necessarily, since a TeV scale for new flavor physics coupling
mainly to third family fermions is not yet ruled out.
\item ``A guaranteed signature of dynamical electroweak
symmetry breaking is some kind of $\rho $-like resonance at a TeV.''

--- Not necessarily, if the new strong interaction at a TeV is itself
broken.  The effects of strong dynamics in that case may not look much
like QCD dynamics.
\item ``Dynamical symmetry breaking destroys the concept of
coupling constant unification.''

--- We will suggest below why the observed values of the
couplings may not be completely accidental, even when new
strong interactions are present.

\end{itemize}

We will be illustrating all the above points in an explicit model. At the
very least, the model will hopefully illustrate why the following two
statements are false.
\begin{itemize}

\item ``Elementary scalar fields must be introduced to build
realistic theories of flavor.''

\item ``Fined-tuned 4-fermion operators must be introduced to
build realistic theories of flavor.''

\end{itemize}
These two statements are given together because the
fine-tuning sometimes invoked in connection with 4-fermion
operators is closely related to the fine-tuning associated with light
elementary scalar fields. It is well known that when fine-tuned
4-fermion operators are arranged to generate small dynamical
fermion masses, they also generate light
composite scalar particles. We clearly want to avoid such
fine-tuning.

\section{Anomalous Scaling}

The key point for the following discussion is the possibility that
the anomalous scaling present in strongly interacting theories can
turn 4-fermion operators into relevant, or nearly relevant,
operators. That is, 4-fermion operators have an effective
dimension close to four instead of six. Such
operators can play the role of Yukawa couplings. They provide an
alternative method for feeding down flavor physics at high scales
into the quark and lepton mass spectrum, without the introduction
of scalar fields.

Relevant 4-fermion operators were found to arise in quenched
QED in ladder approximation.\cite{b2}  But the phenomenon
appears more general, and it is closely related to the tendency for
the scaling dimension of the mass operator $\overline{\psi }\psi $
to become two (anomalous dimension ${\gamma }_{m}=1$) when the
gauge coupling is above some critical value,
$\alpha >{\alpha }_{c}$. Arguments supporting this conclusion have been
made using the operator product expansion,\cite{b3} and using
general properties of the Schwinger-Dyson equation.\cite{b4} If
this is accepted then a large anomalous dimension for
$\overline{\psi }\psi
\overline{\psi }\psi $, perhaps close to two, can also be expected. The
essential requirement is a $\beta $-function which is small over some range
of momentum scales. Nontrivial infrared fixed-points are clearly
of interest in this context.

The important point is that fine-tuning is not required. This is quite
different from the fine-tuning of 4-fermion operators in the
Nambu-Jona-Lasino model which produces an effective
anomalous dimension twice as large: ${\gamma }_{m}=2$. The latter is
related to the existence of light scalar $\overline{\psi }\psi $ bound states.  In
our case 4-fermion operators are relevant at low energies because
the fermions feel a new strong interaction, and not because the
coefficients of the 4-fermion interactions are finely tuned. 
Instead of fine-tuning the physics at one scale to produce desired
results at some much larger distance scale, the anomalous scaling
in our case is arising due to strong interactions acting over a range
of scales.

The operators of interest involve normal quarks and leptons,
and thus the new gauge interactions acting on quarks and leptons at high
energy scales must clearly be broken. These strong interactions must also
be prevented from producing large masses for quarks and leptons. This is
reasonable in a chiral gauge theory in which all masses break gauge
symmetries. The 4-fermion operators which are being enhanced, on the
other hand, need not break any gauge symmetries.

The basic framework will share a few features with extended
technicolor theories.
\begin{itemize}
\item A fermion condensate is responsible for
electroweak symmetry breaking.

\item Four-fermion operators, in the presence of this condensate, produce
all other quark and lepton masses, including the top.
\end{itemize}

The minimal version we will describe differs from basic ETC in the
following ways.

\begin{itemize}
\item There are new strong gauge interactions above a TeV, but they do
not survive below a TeV. There is no unbroken technicolor.

\item We may call the new gauge interactions flavor interactions.
They break down to a remnant flavor symmetry at some scale like 100
or 1000 TeV, and the remnant survives down to a TeV.

\item The third family, as well as a new fourth family, couple to the remnant
flavor interactions, while the light two families couple only to the
flavor interactions broken at the higher scale.

\item The dynamical fourth family quark masses are responsible for
electroweak symmetry breaking. There are no technifermions.

\item The 4-fermion operators responsible for the quark masses
are not of the form
${\overline{H}}_{L}{H}_{R}{\overline{q}}_{R}{q}_{L}$, but are
rather of the form
${\overline{q}}_{L}{H}_{R}{\overline{H}}_{L}{q}_{R}$ where \(H\)
denotes a fourth family quark. (These operators cannot arise by explicit gauge
boson exchange and must therefore originate dynamically.)

\item The 4-fermion operator responsible for the $t$ mass is composed of
third and fourth family quarks, and is strongly enhanced by the remnant
flavor interactions. This particular operator may be close
to being a relevant 4-fermion operator, as discussed above.

\item  The dynamical fourth family quark masses are not in a singlet
channel with respect to the remnant flavor interactions.
\end{itemize}

\section{A Model of Flavor Physics}

We will illustrate these ideas by considering the most minimal
flavor gauge symmetry we can imagine:\cite{b5}
\begin{equation}
U(1{)}_{V}\times SU{(2)}_{V},
\end{equation}
such that the complete
quark content of the theory is given by
\begin{equation}
Q\;:\;(+,2)\;\;\;\;\;\;\;\;\;\;\underline{Q}\;:\;(-,\overline{2}).
\end{equation}
Each \(Q\) field also
transforms under the standard model gauge group as a normal quark, and
thus there are enough fields here to describe the quarks of four
families.\footnote{To actually make this a chiral gauge group a
further \(U(1)\) interaction may be added, or alternatively some
relevant 4-fermion operator may play an equivalent role in
making the dynamics chiral.\cite{b5}} The flavor symmetry
breaks at a scale
$\Lambda
\approx$ 100 to 1000 TeV to $U(1{)}_{X}$, which is an unbroken
combination of
$U(1{)}_{V}$ and the diagonal generator of $SU(2{)}_{V}$.
Then the $X$ boson only couples to the third and fourth family
quarks ${Q}_{1}$ and
${\underline{Q}}_{1}$,
and not to the light quarks
${Q}_{2}$ and ${\underline{Q}}_{2}$.

Flavor changing neutral currents involving the light two families are
suppressed by inverse powers of \(\Lambda\). We shall argue below that
most of the mass mixing between families occurs in the up-sector. This
combined with the fact that flavor interactions do not induce transitions
between the
${Q}_{2}$ and ${\underline{Q}}_{2}$ fields will suppress further the
most problematical FCNC effect, \(K\)--\(\overline{K}\) mixing.

It is important to note that none of the fields
${Q}_{1}$, ${\underline{Q}}_{1}$,
${Q}_{2}$, ${\underline{Q}}_{2}$ are mass eigenstates. In fact the
fourth family quark masses will correspond to
\begin{equation}
{\overline{\underline{Q}}}_{1L}{Q}_{1R}+{\rm h.c.}
\label{bb}
\end{equation}
As we have said,
this is not a singlet under the remnant flavor symmetry
$U(1{)}_{X}$, and it can only form once the $U(1{)}_{X}$
breaks (we discuss this breakdown below). Although the
\(U(1)_X\) is strong it is unclear whether it, along with possible
4-fermion interactions, is sufficient to produce the mass in (\ref{bb}).
We will continue our discussion of the minimal flavor interactions
and leave open for now the question of whether additional
interactions are required. In the appendix we describe how an
enlarged color interaction can help to produce these fourth family
quark masses.

We now consider a set of
operators which will feed mass down from the fourth family
quarks to the second and third family quarks.
\begin{eqnarray*}
({\overline{Q}}_{L}{D}_{R})\epsilon ({\overline{\underline{Q}}}_{L}{\underline{U}}_{R})&&{\cal B}\\
({\overline{Q}}_{L}{U}_{R})\epsilon({\overline{\underline{Q}}}_{L}{\underline{D}}_{R})&&\tilde{{\cal B}}\\
({\overline{Q}}_{L}{D}_{R})\epsilon ({\overline{\underline{Q}}}_{L}{\hat\epsilon} {U}_{R})&&{\cal C}\\
({\overline{Q}}_{L}{U}_{R}) \epsilon({\overline{\underline{Q}}}_{L}{\hat\epsilon}{D}_{R})&&\tilde{{\cal C}}\\
({\overline{\underline{Q}}}_{L}{\hat\epsilon} {D}_{R})\epsilon ({\overline{\underline{Q}}}_{L}{\underline{U}}_{R})&&{\cal D}\\
({\overline{\underline{Q}}}_{L}{\hat\epsilon}{U}_{R})\epsilon ({\overline{\underline{Q}}}_{L}{\underline{D}}_{R})&&\tilde{{\cal D}}\\
({\overline{\underline{Q}}}_{L}{\hat\epsilon}{D}_{R})\epsilon ({\overline{\underline{Q}}}_{L}{\hat\epsilon} {U}_{R})&&{\cal E}
\end{eqnarray*}
The $\epsilon$ contracts $SU(2)_L$ indices while the \({\hat\epsilon}\) 
contracts \(SU(2)_V\) indices. Note that all these operators are
$SU(2)_V$ singlets, i.e. in an attractive channel with respect to
$SU(2{)}_{V}$ interactions. Thus although these operators are not
generated by an explicit $SU(2{)}_{V}$ gauge boson exchange, they
could be expected to be produced dynamically by strong
$SU(2{)}_{V}$ interactions. There may well be other operators
generated, but these are the operators of most interest for the
generation of mass.

These operators have differing flavor structure, as
indicated by the sprinkling of the \({\hat\epsilon}\)'s, and this
results in nontrivial mass matrices, to which we now turn. We first
consider \(2\times2\) blocks of the full \(4\times4\) mass matrices.
The
$c$--$t$ submatrix, in terms of the original fields, is
\begin{equation}
\left({\begin{array}{cc}
{\overline{\underline{U}}}_{2L}{U}_{2R}&{\overline{\underline{U}}}_{2L}{\underline{U}}_{1R}\\
{\overline{U}}_{1L}{U}_{2R}&{\overline{U}}_{1L}{\underline{U}}_{1R}
\end{array}
}\right)
.\end{equation}
The operators as labeled above contribute in the following way.
\begin{equation}
\left({\begin{array}{cc}
{\cal E}&{\cal D}\\
{\cal C}&{\cal B}
\end{array}
}\right)\;\;\;\;\;\Rightarrow \;\;\;\;\;c{\rm \;and\;}t{\rm \;masses}
\end{equation}
Note that these elements are being fed down from the $b'$ mass. Similarly
the $t'$ mass is feeding down to the $s$ and $b$ masses as
follows.
\begin{equation}
\left({\begin{array}{cc}
{\cal E}&\tilde{{\cal D}}\\
\tilde{{\cal C}}&\tilde{{\cal B}}
\end{array}
}\right)\;\;\;\;\;\Rightarrow \;\;\;\;\;s{\rm \;and\;}b{\rm \;masses}
\end{equation}

Not only do we get a diverse set of operators with different structures
emerging from a fairly simple starting point, but each operator
contributes to a different element of a mass matrix. We see only one
common element in the up- and down-type matrices, and thus
\(t\)--\(b\) mass splitting is permitted. In addition, the
different off-diagonal elements imply nontrivial mass mixing
between the 2nd and 3rd families. But the main point about the
way masses are being generated here is the way the dynamics
generates mass hierarchies. For example the generation of the
${\cal B}$ and
\(\tilde{{\cal B}}\) operators are favored because among the
operators we have listed, only they are singlets under
$U(1{)}_{V}$. More importantly, the operators experience
different amounts of anomalous scaling, due to the remnant
$U(1{)}_{X}$ interaction, as they are run down from scale $\Lambda $
where they are created, to a TeV.  We expect that the
${\cal B}$ and \(\tilde{{\cal B}}\) operators are enhanced the most while the
${\cal E}$ operator is enhanced the least. In particular at one-loop, the
$U(1{)}_{X}$ corrections enhance the ${\cal B}$ and \(\tilde{{\cal B}}\)
operators, they cancel out for the ${\cal C}$ and ${\cal D}$ operators, and
they resist the ${\cal E}$ operator. The following hierarchy follows.
\begin{equation}
{\cal B},\tilde{{\cal B}}>{\cal C,D},\tilde{{\cal C}},
\tilde{{\cal D}}>{\cal E}
\end{equation}

What we have not explained is the origin of the $t$--$b$ mass splitting.
That must be due to the larger size of the \(({\cal B},{\cal
C},{\cal D})\) operators compared to the \((\tilde{{\cal
B}},\tilde{{\cal C}},\tilde{{\cal D}})\) operators, due to physics
at the high scale. This could for example be related to a dynamical
breakdown of
$SU(2{)}_{R}$ in an otherwise left-right symmetric theory. But
whatever the origin of the isospin breaking, it is manifested at a
TeV in various 4-fermion operators. We find that the most
dominant of these operators do not feed directly into $\delta \rho
$; in fact four insertions of the dominant ${\cal B}$ operator is
necessary to produce a contribution to $\delta \rho
$.\cite{b5a,b5} This produces a suppression factor of order
$(m_t/m_{t'})^4$. It may be checked that other possible
operators contributing to
\(\delta\rho\) are not of the form which is strongly enhanced by
the \(U(1)_X\) scaling effect. This applies to the
$\overline{H}H\overline{H}H$ operators mentioned in the
introduction, which in the present case corresponds to operators
with four fourth-family quarks. In particular it is not possible to write these
operators as a product of Lorentz scalars which are neutral under
\(U(1)_X\). We thus see how the dynamics protects \(\delta\rho\)
from the isospin breaking physics responsible for the
large \(t\) mass.

We return to the question of the breakdown of $U(1{)}_{X}$,
where we note that this symmetry is broken by the ${\cal C}$,
${\cal D}$ and ${\cal E}$ operators. Since these operators are
arising dynamically, the implication is a dynamical contribution
to the
$X$ boson mass.  What contribution does a 4-fermion condensate
give to a gauge boson mass? Using naive dimensional
analysis\cite{b6} (which tries to keep track of 4$\pi$ factors in
loops) it has been argued\cite{b5} that it is natural to expect a
contribution to the mass over coupling of the gauge boson,
$M/g$, an order of magnitude or two less than the natural scale of
the operator. In other words, 4-fermion condensates seem to
produce a softer breaking of gauge symmetries than do 2-fermion
condensates.\footnote{This would also imply that 4-fermion
condensates are less resisted by repulsive gauge interactions than
2-fermion condensates.} In this way we find that
\(M/g\) for the \(X\)-boson could be as low as a TeV or so.

\section{The Role of Leptons}

We may introduce leptons in the same way as quarks, such that
under
$U(1{)}_{V}\times SU(2{)}_{V}$
\begin{eqnarray*}
{Q}\;{\rm and}\;L&{\rm transform\;as}&(+,2),\\
\underline{Q}\;{\rm and}\;\underline{L}&{\rm
transform\;as}&(-,\overline{2}).
\end{eqnarray*}
Operators of interest involving leptons are as listed. (We do not
consider operators involving right-handed neutrinos, since the
latter may not exist in the theory at a TeV.)
\begin{eqnarray}
({\overline{\underline{L}}}_{L}{\underline{E}}_{R})\epsilon ({\overline{Q}}_{L}{U}_{R})&&{\cal
F}\nonumber\\ ({\overline{L}}_{L}{E}_{R})\epsilon
({\overline{\underline{Q}}}_{L}{\underline{U}}_{R})&&\nonumber\\
({\overline{L}}_{L}{E}_{R})\epsilon
({\overline{Q}}_{L}{U}_{R})&&\nonumber\\
({\overline{\underline{L}}}_{L}{\underline{E}}_{R})\epsilon
({\overline{\underline{Q}}}_{L}{\underline{U}}_{R})&&{\cal G}
\label{r2}\end{eqnarray}
These operators respect all gauge symmetries, like the ${\cal B}$
operators. And as we shall show, the ${\cal F}$ and ${\cal G}$
operators make additional contributions to the quark masses if the
$\tau'$ mass corresponds to
${\overline{\underline{E}}}_{1L}{\underline{E}}_{1R}$. We
therefore assume that that is the case.

It turns out that the operators in (\ref{r2}), along with our previous
operators, are sufficient to break all global symmetries. This
removes the necessity of some kind of quark-lepton unification at
relatively low scales, and thus allows the standard
$SU(3)\times SU(2)\times U(1)$ (or left-right symmetric version)
to survive to high scales. Since the fermions of our model come in
standard model families, the relative running of the $SU(3)\times
SU(2)\times U(1)$ couplings remains the same to lowest order in
these couplings, to {\em all} orders in the new strong interactions.
Thus there remains the tendency for the three couplings to
become closer at high energies. In other words, if there is some
form of coupling constant unification then the observed values of
the three couplings are not completely accidental in this picture
in spite of new strong interactions.

We may now display the full up- and down-type mass matrices.
\begin{equation}
{M}_{u}=\left({\begin{array}{cccc}
0&{{\cal F}}_{2}&0&0\\
{{\cal G}}_{2}&{\cal E}&{\cal D}&0\\
0&{\cal C}&{\cal B}&{{\cal F}}_{1}\\
0&0&{{\cal G}}_{1}&{\cal A}
\end{array}
}\right)
\end{equation}
\begin{equation}
{M}_{d}=\left({\begin{array}{cccc}
{\cal H}&0&{\cal I}&0\\
0&{\cal E}&\tilde{{\cal D}}&0\\
{\cal I}&\tilde{{\cal C}}&\tilde{{\cal B}}&0\\
0&0&0&{\cal A}
\end{array}
}\right)
\end{equation}
The \({\cal A}\) entry is the dynamical fourth-family quark mass
in (\ref{bb}). The other new entries in the up-type matrix are
due to the quark-lepton operators, which are feeding mass down
from the
$\tau'$. Note that the matrix is in general not symmetric. We are
postulating that most of the KM mixing angles have their origin
in the up-mass matrix. In particular Cabibbo mixing is due
mostly to the ${\cal F}_2$ entry, while the small size of the \(u\)
mass is due to the smaller ${\cal G}_2$ entry.

In the down-type
matrix, the
${\cal H}$ entry is due to an operator similar to the
${\cal E}$ operator, but which  is feeding mass down from the
$t$ rather than the \(t'\). (The ${\cal H}$ entry in the up-type
matrix is negligible since it is feeding down from the \(b\).)  This
leads to the mass relation
\begin{equation} {\frac{{m}_{d}}{{m}_{t}}}\approx
{\frac{{m}_{s}}{{m}_{t'}}}.
\end{equation}
It is rather novel to find that the $u$ mass is being fed down from the
$\tau'$, while the $d$ is being fed down from the $t$. The ${\cal I}$
operator, and perhaps the ${\cal E}$ and ${\cal H}$ operators as well,
may be arising through loops involving other operators we have listed.
It may be shown that mass matrices of the above form have sufficient
structure to produce realistic quark masses and mixings.

In the lepton sector we are able to identify 4-fermion operators which
could give mass to the charged leptons. The small mass of left-handed
neutrinos is probably associated with a large mass for right-handed
neutrinos. But we are unable to make predictions for the three light
neutrino masses other than that they are unlikely to be much
smaller than an eV.

\section{Signatures}

We return to the point that the ${\cal E}$ entry is the
same for the up- and down-type matrices, which is due to the fact
that the
${\cal E}$ operator is intrinsically isospin conserving. To avoid
unnatural cancellation in the
\(b\)--\(s\) mass matrix, this entry should be of order the
$s$ mass. This means that the $c$ mass must be due mostly to \(c\)--\(t\)
mixing induced by the
${\cal C}$ and ${\cal D}$ entries. These entries are thus of order
$\sqrt{{m}_{t}{m}_{c}}$, and this in turn has an interesting
phenomenological consequence.

If we take one of our operators and close off the $t'$ or $b'$ lines
we get a contribution to the quark mass matrix. If we also attach a
photon or $Z$ to the heavy quark loop then we generate an
anomalous magnetic-moment-type coupling; in particular we can
get the flavor-changing coupling
\begin{equation}
\overline{t}{\sigma }_{\mu \nu }{q}^{\nu }(1,{\gamma
}_{5})c\;({A}^{\mu },{Z}^{\mu })+{\rm h.c.}
\end{equation}
Since the size of this operator is closely related to the corresponding
off-diagonal mass elements, we estimate that its coefficient is of
order
\begin{equation} {\frac{e{m}_{c}{m}_{t}}{{\lambda }^{2}}}
\end{equation}
where $\lambda \approx $TeV. This is larger than the
conventional ETC estimate, since ETC mass
operators do not directly generate these couplings. It is also larger
than estimates based on multi-Higgs models, since in that case the
new coupling requires an extra loop compared to the tree-level 
mass.

This coupling yields
\begin{eqnarray}
{R}^{ct}&\equiv &{\frac{\sigma ({e}^{+}{e}^{-}\rightarrow t\overline{c}+\overline{t}c)}{\sigma ({e}^{+}{e}^{-}\rightarrow \gamma \rightarrow {\mu }^{+}{\mu }^{-})}}\\
&\propto &{\frac{(s+2{m}_{t}^{2})(s-{m}_{t}^{2}{)}^{2}}{3{s}^{2}{m}_{t}^{2}}}(1-{\rm cos}(\theta
{)}^{2})
\end{eqnarray}
with an magnitude which is just barely detectable at LEP2, and easily
detectable at 500 GeV. Chromomagnetic moments may also be
considered, and in that case the flavor-diagonal anomalous coupling of a
gluon to top quarks may be of interest to the production of top quarks in
hadron colliders.

Generic to this picture are the remnant flavor interactions, which
implies at least one new massive gauge boson coupling to the
third family with a mass as low as a TeV.\cite{b6a} In the
minimal model this is the \(X\)-boson, and we find that it mixes
with the
$Z$. As we have described it, the \(X\)-boson has axial couplings
to the \(t\) and \(b\) quarks and vector couplings to the
\(\tau\).\footnote{The latter is a byproduct of the choice
${\overline{\underline{E}}}_{1L}{\underline{E}}_{1R}$ for
the $\tau'$ mass} The mixing then results in the following wide
range of fractional shifts in various electroweak parameters,
which occur in the ratios,
\[{{\cal A}}_{\tau
}\;\;:\;\;R_b\;\;:\;\;\Gamma_{\nu_\tau}\;\;:\;\;{\cal
A}_b\;\;:\;\;\Gamma_\tau\]
\[+20\;:\;+2.0\;:\;-1.5\;:\;-0.5\;:\;+0.2\]
The magnitude of the shift of the $Z$ coupling to a third
family fermion is
\begin{equation}
\delta{g}_{Z}={\frac{e}{sc}}{f}_{t}^{2}{\frac{{g}_{X}^{2}}{{M}_{X}^{2}}}
.\end{equation}
$f_{t}\approx 60\;{\rm GeV}$ is determined from the $t$ loop with
an effective TeV cutoff supplied by the momentum dependence of the
$t$ mass. For example, an observed 2\% shift in $R_{b}$ or a 20\% shift
in ${\cal A}_{\tau }$ would imply that
\begin{equation}
{\frac{M_{X}}{g_{X}}}\approx 1\;{\rm TeV}.
\end{equation}
What is perhaps most surprising is that such large anomalies have not yet
been completely ruled out.\footnote{We note in particular the current
anomaly in ${\cal A}_{\tau }/{\cal A}_{e,\mu}$ extracted from the
forward-backward lepton asymmetries at LEP.\cite{b7}}

We note in passing that the
new flavor physics is not associated (at least in the minimal
$U(1)_{X}$ picture presented here) with any kind of new confining
interaction. This new flavor physics could be quite different from QCD-like
dynamics and the implied $\rho$-like resonance typically expected in theories
of dynamical symmetry breaking.

And finally, this picture leads a fourth left-handed neutrino,
with a dynamical mass expected to be comparable to the
other fourth-family members. Assuming that the right-handed
neutrino is either absent or much heavier, we are left to consider
a large Majorana mass for the left-handed neutrino. This might
seem to be a disaster for electroweak corrections, and for this reason
the effects of
Majorana masses are normally only considered for right-handed
neutrinos.\cite{b7a} But this expectation is not correct, at least for some range
of masses. In Ref. \cite{b8} we considered the contributions to
$S$, $T$, and $U$ from fourth family leptons $({\nu }_{L},{\tau
}^{\prime })=(N,E)$ (we omit the underlines) for a range of masses
${m}_{N}$ and ${m}_{E}$. The result for
$T$ depends on the effective cutoff in the neutrino loop (again
supplied by the momentum dependence of the dynamical mass) and we
have used $\Lambda = 1.5m_N$ and $\Lambda = 2m_N$. From
Figs. (1) and (2) we see ranges of masses for which the new
contributions to
$T$ are negative with reasonable size, while the new
contributions to $S$ and $U$ are simultaneously small. We
therefore have a source of negative
$T$ within the dynamical symmetry breaking context. This may
be useful, given the fact that other new contributions to $T$ in
this context are typically positive.

In conclusion, we have described a picture in which flavor physics originates
at a scale of order 100 or 1000 TeV. We have emphasized the role that
anomalous scaling of 4-fermion operators can have in transforming
the flavor physics at this scale into the complicated pattern of observed
fermion masses.

\section*{Appendix}
We present here an extension of the minimal model which may have more
appealing dynamics. In particular we replace the
$U(1)_X$ with
$SU(3)_{X}$. Each third and fourth
family fermion is now one member of an
$SU(3)_X$ triplet. We also expand color into a direct product
of $SU(3)$'s, such that under the gauge symmetry
\begin{equation}
SU(3)_{X}\times SU(3)_{C1}\times SU(3)_{C2},
\label{r1}\end{equation}
the third and fourth family quarks are members of the following
representations.
\begin{eqnarray}
Q_{L}\;\;\;(3,3,1)\;\;\;\;\;\;\;\underline{Q}_{L}\;\;\;(\overline{3},1,3)&&\\
Q_{R}\;\;\;(3,1,3)\;\;\;\;\;\;\;\underline{Q}_{R}\;\;\;(\overline{3},3,1)&&
\end{eqnarray}
The symmetry in (\ref{r1}) is not a symmetry of the $\cal  C$ and $\cal
D$ operators, and as before we assume that the gauge bosons
corresponding to the broken generators have masses in the TeV range. The
symmetry left unbroken by the $\cal  C$ and $\cal D$ operators is
$SU(2)_{M}\times SU(3)_{C}$, where the former is a ``metacolor'' and
the latter is standard color. We may assume that
$SU(3)_{C1}$ is strong enough to help
induce the $SU(3)_{C1}$ conserving and $SU(3)_{X}$ violating
condensate
$\left\langle{\overline{Q}_{1L}\underline{Q}_{1R}}\right\rangle$. 
The condensate lies in the {\bf
6} of $SU(3)_X$, and it can be transformed to the given `11' form. The
result is the fourth family quark masses in (\ref{bb}), and the main point is
that there is now an obviously attractive gauge interaction in this channel.

Since $SU(3)_{2C}$ is likely weaker than $SU(3)_{1C}$ in order to get
the correct \(\alpha_C\), the `broken-color' interactions will end up coupling
more strongly to the second and fourth families than to the first and third. As
for metacolor, we assume it grows strong somewhat below a TeV and that it
confines the two families of metafermions. If metafermions become massive
then they make an unwelcome contribution to the $S$ parameter. On the other
hand their chiral symmetries are already explicitly broken by the
broken gauge interactions, and thus the confined metafermions could
remain massless due to unbroken discrete chiral symmetries
without constraints from chiral anomalies.\cite{b8a}

This whole structure can easily be incorporated into the larger
gauge group present at 100 or 1000 TeV. The $SU(3)_X$
becomes embedded into
$SU(4)_V$, which acts also on the two light families. All the
4-fermion operators we have considered may be chosen to transform as
singlets under an $SU(2)$ subgroup of $SU(4)_V$, disjoint from the
unbroken $SU(2)_M$ subgroup. The operators are singlets under this
$SU(2)$ subgroup in the same way they were singlets under the $SU(2)_V$
of the minimal model. It may be checked that they are therefore in attractive
channels with respect to
$SU(4)_V$ (even though they do not respect $SU(4)_V$), just as they were
in attractive channels with respect to
$SU(2)_V$. Finally we note that the fermion content of the model implies that
$SU(4)_V$ may be a candidate for a nontrivial infrared fixed point according
to the analysis in Ref. \cite{b9}.

\section*{Acknowledgments} 
I thank the organizers of the Strong Coupling Gauge Theories 96
meeting in Nagoya for a stimulating conference and for their support.
This research was supported in part by the Natural Sciences and
Engineering Research Council of Canada.

\newpage
\begin{center}
\includegraphics{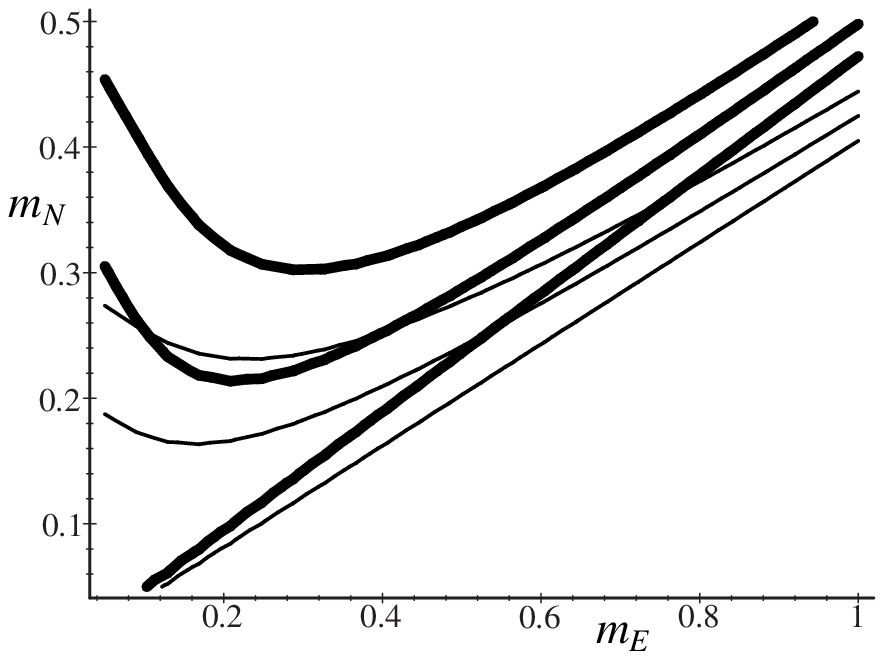}
\end{center}
\noindent Figure 1: Lines of constant $T$ as a function of the $N$ and $E$
masses in TeV. Thick and thin lines are for $\Lambda = 1.5m_N$ and
$\Lambda = 2m_N$ respectively. In each case, from top to bottom,
$T=-2,-1,0$.
\vspace{7ex}
\begin{center}
\includegraphics{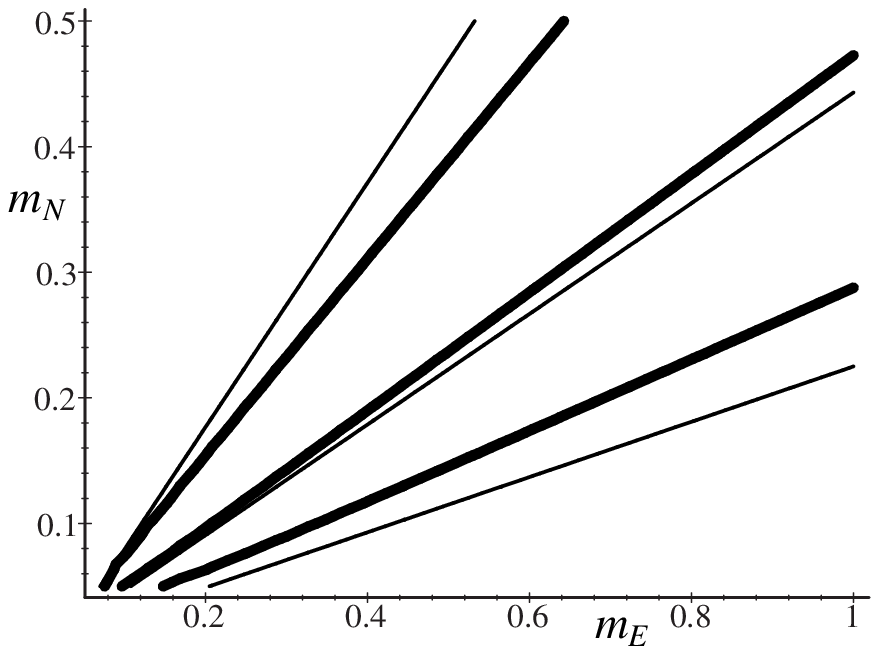}
\end{center}
\noindent Figure 2: Thick and thin lines are lines of constant $S$ and $U$
respectively as a function of the $N$ and $E$ masses in TeV. From top to
bottom in each case $S=1/6\pi,0,-1/6\pi$ and $U=-1/12\pi,0,1/6\pi$.

\end{document}